# Direct and Indirect Effects of Sequential Treatments

**Vanessa Didelez**
Department of Statistical Science
University College London, UK
vanessa@stats.ucl.ac.uk

**A. Philip Dawid**
Department of Statistical Science
University College London, UK
dawid@stats.ucl.ac.uk

**Sara Geneletti**
School of Medicine
Imperial College London, UK
s.geneletti@imperial.ac.uk

## Abstract

In this paper we review the notion of direct and indirect causal effect as introduced by Pearl (2001). We show how it can be formulated without counterfactuals, using regime indicators instead. This allows to consider the natural (in)direct effect as a special case of sequential treatments discussed by Dawid & Didelez (2005) which immediately yields conditions for identifiability as well as a graphical way of checking identifiability.

## 1 INTRODUCTION

We consider the question of causal inference as a question of inference across different *regimes*: what we would like to know is the behavior of a system (or just the response variable) under a regime in which we intervene in certain variables in certain ways, e.g. by fixing them to pre–specified values, while the information available to us, i.e. the data, is typically generated under an *observational* regime, in which no interventions take place. Causal inference is possible, if these two regimes can be 'linked' in some sense. A corresponding formal framework is based on decision variables or regime indicators and influence diagrams (Dawid, 2002).

In Dawid & Didelez (2005) this approach is applied to address the question of identifiability of sequential treatments as briefly reviewed in Section 2.4 below. Here we consider, for simplicity, the situation of only two sequential treatments, one treatment $X_1$ given at an earlier point in time, a second treatment $X_2$ given later possibly depending on some covariate $V$ that has been observed after $X_1$ but before $X_2$ and a final outcome variable $Y$. It is known due to the seminal work by Robins (1986, 1987) that the causal effect of $X_1$ and $X_2$ on $Y$ has carefully to take into account $V$ which is a potential mediating variable for the effect of $X_1$ but a confounder for the effect of $X_2$. A suitable adjustment can be carried out using the g–formula. Graphical criteria for the identifiability of unconditional sequential interventions via the g–formula are derived in Pearl & Robins (1995) and for dynamic, i.e. conditional, interventions in Dawid & Didelez (2005).

In this context, the notion of direct effect adapted from Pearl (2001) can be described as the effect of changing $X_1$ while $X_2$ is kept 'constant' in some sense, e.g. by controlling its conditional distribution. Hence, formally, this is similar to a sequential treatment setting. But it is also different because for the *natural* direct effect the conditional distribution that $X_2$ is generated from is not necessarily known.

The outline of the paper is as follows. We first present in Section 2 some prerequisites to explicate the framework of causal inference as inference across regimes. In Section 3 we define different notions of direct and indirect effects in terms of interventions. Their identifiability in different data situations is addressed in Section 4 with particular emphasis on the identification from observational data. This is followed by a discussion in Section 5.

## 2 PRELIMINARIES

We assume throughout that all variables are discrete (for the continuous case one essentially just needs to replace pmfs by pdfs and sums by integrals). We also assume that the conditional independencies among the variables of interest can be represented in a DAG $G$, i.e. the Markov properties of $G$ are satisfied by the joint distribution.

### 2.1 REGIME INDICATORS

As we consider causal inference as inference across regimes, our approach uses regime *indicators* (Dawid, 2002; Dawid & Didelez, 2005), an idea going back to Pearl (1993). The regime indicator for an intervention in a variable $X$ is denoted by $\sigma_X$ and can take values in a set $\mathcal{S} \cup \emptyset$ of strategies. In the context of a DAG, these strategies define the conditional distribution of $X$ given its graph parents. Let $p(x|\text{pa}_X; \sigma_X = s)$ denote the conditional pmf of $X$ under regime $s \in \mathcal{S} \cup \emptyset$. The following examples illustrate the different types of regimes that are of interest here.

*Observational regime:* The observational or idle regime $\sigma_X = \emptyset$ means no intervention takes place, i.e. the conditional distribution is just the one that generates the observational data

$$p(x|\text{pa}_X; \sigma_X = \emptyset) := \text{observational}.$$

The above, by definition, can be estimated from data if $\text{pa}_X$ are observed. Any parameter that can be expressed in terms of observational distributions is (non–parametrically) identifiable.

*Atomic intervention:* The strategy of setting $X$ to a fixed value $x^*$, denoted by $do(x^*)$ in Pearl (2000), is here symbolized by $\sigma_X = s_{x^*}$ such that

$$p(x|\text{pa}_X; \sigma_X = s_{x^*}) := \delta(x, x^*),$$

where $\delta(a, b)$ is one if $a = b$ and zero otherwise. Hence $X \perp\!\!\!\perp \text{pa}_X | \sigma_X = s_{x^*}$, graphically implying that the arrows entering $X$ are cut off. The strategy $\sigma_X = s_{x^*}$ describes an intervention that sets $X$ to $x^*$ regardless of the values of other variables, e.g. choosing the dosage of a treatment regardless of other aspects or information about a patient. Due to the deterministic relation between $\sigma_X = s_{x^*}$ and $X$ any other variable is independent of $X$ given $\sigma_X = s_{x^*}$, e.g. if $Y$ is the response variable then $p(y|x; \sigma_X = s_{x^*}) = p(y; \sigma_X = s_{x^*})$.

*Conditional intervention:* More generally we may want the value of $X$ to depend on the previously observed variables, i.e. on $\text{pa}_X$. Let $a(\cdot)$ be a pre–specified function indicating the values that we want to set $X$ to once we have seen $\text{pa}_X$, then

$$p(x|\text{pa}_X; \sigma_X = s_{a(\text{pa}_X)}) := \delta(x, a(\text{pa}_X)).$$

Given we follow such a conditional strategy, $X$ is obviously not conditionally independent of its parents anymore. This is the case e.g. when the dosage of a treatment is chosen conditional on the sex and age of a patient according to a pre–specified plan and the overall effect of such a conditional strategy is to be evaluated. In case of interactions between $X$ and $\text{pa}_X$ in their effect on $Y$ it is particularly sensible to consider such conditional interventions as an optimal strategy will typically have to be conditional.

*Random (conditional) intervention:* Even more generally, we may not just want to fix $X$ at a specific value, but let it take on a random value according to some distribution possibly depending on $\text{pa}_X$. Such a strategy is denoted by $\sigma_X = d_{\text{pa}_X}$ with

$$p(x|\text{pa}_X; \sigma_X = d_{\text{pa}_X}) := \tilde{p}(x|\text{pa}_X),$$

where $\tilde{p}(x|\text{pa}_X)$ is a pmf in $x$ possibly depending on some of $\text{pa}_X$. A random strategy can describe the situation where treatment is randomized possibly within strata of covariates, and hence represents what we expect to see in a designed experiment. In the context of (in)direct causal effects we need to represent conditional random interventions, such that $X$ is generated from its observational distribution with at least one of its parents set to a fixed value. Note that this distribution may not be known, e.g. in a placebo study the psychological effect of a subject thinking that he or she is being treated is not itself fixed at some level, but it is made sure to arise in the same way for everyone because every subject receives an identical tablet.

Regime indicators correspond to decision variables and do not have marginal distributions. They can be regarded as indexing the considered conditional distributions. Therefore, these as well as all (conditional) independence statements made have to specify under which regime they hold. However, for notational ease we adopt the convention that when no regime indicator $\sigma_X$ is specified we mean the observational regime, i.e. $\sigma_X = \emptyset$. Furthermore, statements like $Y \perp\!\!\!\perp \sigma_X | X$ mean that $p(y|x; \sigma_X = s) = p(y|x; \sigma_X = s')$ for any $s, s' \in \mathcal{S} \cup \emptyset$, $s \neq s'$ (cf. Dawid, 2002).

## 2.2 INFLUENCE DIAGRAMS

In general, one should draw a separate DAG for each regime under consideration as there is a priori no reason to assume that the joint distribution of the considered variables is the same under a variety of interventions and manipulations. However, as we want to address inference across regimes, we need to express the conditions allowing such inference in a common influence diagram where those conditional specifications that do not involve regime indicators are assumed to remain the same under different regimes (Dawid, 2002). Such an influence diagram is a DAG that includes regime indicators as nodes of their own, drawn with a box in order to make clear that these are decision nodes, like for instance in Figure 1.

Further we want any indicator $\sigma_X$ to only have an arrow into $X$, i.e. only into the node that it intervenes in. The resulting graph implies certain conditional independencies, e.g. all descendants of $X$ are conditionally independent of $\sigma_X$ given $X$ and their parent set. In order for these independencies to hold in a given data situation one typically has to add further variables in the graph that might not be observable, like for instance in Figure 4.

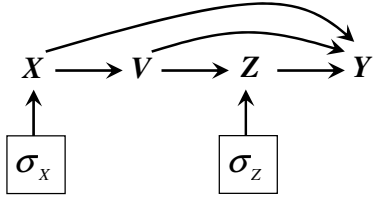

Figure 1: Influence diagram with indicators for interventions in $X$ and $Z$.

Finally note that the parent set pa$_X$ of a node $X$ may vary under different regimes, e.g. as mentioned above $\sigma_X = s_x$ cuts off all arrows into $X$ whereas a conditional intervention might add parents as compared to $\sigma_X = \emptyset$ by making the value of $X$ depend on variables that observationally it does not depend on. Hence, in the influence diagram the parent set is the union of the parent sets under all considered regimes. When conditional independence statements like $Y \perp\!\!\!\perp \sigma_X | (X, Z; \sigma_Z = s)$ are verified graphically (using the moralization or d–separation criterion), the arrows into $Z$ that are 'cut off' by $\sigma_Z = s$ should be removed, like e.g. in the influence diagram of Figure 4 modified in Figure 5.

## 2.3 CAUSAL EFFECT

The main ingredient in any of the different notions of causal effects that we present below is the *intervention distribution* $p(y; \sigma_X = s)$ of the response $Y$ for some regime $s \in \mathcal{S}$ applied to $X$ (or more generally to some of the other variables). The *causal effect* of $X$ on $Y$ is typically taken to be some contrast between intervention distributions for two different strategies, $\sigma_X = s_1$ and $\sigma_X = s_0$, $s_1 \neq s_0 \in \mathcal{S}$, say. The *average total effect* of setting $X$ to $x$ as compared to $x^*$

$$ACE(x, x^*; Y) := E(Y; \sigma_X = s_x) - E(Y; \sigma_X = s_{x^*}),$$

where $x^*$ is often a baseline value. However, we will mainly focus on the intervention distribution $p(y; \sigma_X = s)$ itself; if this can be identified so can any contrast. It has to be kept in mind, though, that specific parameters, e.g. the expectation, of the intervention distribution may be identifiable under different conditions, typically referring to particular parameterizations, even when the whole distribution is not.

Conditions for the identifiability of the average causal effect are well known and can be put into graphical terms, e.g. as back–door and front–door criteria (Pearl, 1995). With our notation, the former states that if a set of covariates $C$ can be found, such that $C \perp\!\!\!\perp \sigma_X$ (i.e. $C$ is no descendant of $X$) and $Y \perp\!\!\!\perp \sigma_X | (X, C)$ (i.e. every back–door path is blocked by $C$) the intervention distribution calculates as

$$p(y; \sigma_X = s_x) = \sum_c p(y|c, x) p(c)$$

and

$$ACE(x, x^*; Y) = \sum_c [E(Y|c, x) - E(Y|c, x^*)] p(c),$$

(omitting the condition $\sigma_X = \emptyset$). In case of a conditional and random intervention, i.e. if $X$ is drawn from a distribution $\tilde{p}(x|c)$ specified by the strategy $d_C$ we have the intervention distribution

$$p(y; \sigma_X = d_C) = \sum_{x,c} p(y|c, x) \tilde{p}(x|c) p(c).$$

## 2.4 SEQUENTIAL TREATMENTS

In a sequential decision problem more than one intervention take place, e.g. we may want to intervene in $X_1, \ldots, X_K$. This has been considered by Pearl & Robins (1995) and within the framework of regime indicators by Dawid & Didelez (2005).

To briefly summarize the results relevant for the present article, we need the following notations. Let $\sigma_1, \ldots, \sigma_K$ denote the individual regime indicators for interventions in $X_1, \ldots, X_K$. Let further $L_1, \ldots, L_K$ denote observed covariates, where $(L_1, \ldots, L_k)$ are non–descendants of $X_k$. Further $\bar{X}_k = (X_1, \ldots, X_k)$ denotes the past and $\bar{X}^k = (X_k, \ldots, X_K)$ the future. An intervention in $X_k$ can be conditional on $\bar{X}_{k-1}$ and $\bar{L}_k$ as these are assumed to be observed prior in time.

Then, under conditions addressed below, we can identify the intervention distribution of $Y$ as

$$p(y; \bar{\sigma}_K = \bar{s}_K) = \sum_{\bar{x}_K, \bar{l}_K} p(y|\bar{x}_K, \bar{l}_K)$$
$$\times \prod_{k=1}^{K} p(x_k|\bar{x}_{k-1}, \bar{l}_k; \sigma_k = s_k) p(l_k|\bar{l}_{k-1}, \bar{x}_{k-1}) \quad (1)$$

where $p(x_k|\bar{x}_{k-1}, \bar{l}_k; \sigma_k = s_k)$ are determined by the chosen regime and the other quantities can be estimated from observational data. The above is known as the g–formula (Robins, 1986, 1987; Pearl & Robins, 1995). For (1) to be valid the following is a sufficient condition

$$L_k \perp\!\!\!\perp \bar{\sigma}_K \mid (\bar{L}_{k-1}, \bar{X}_{k-1}) \text{ and } Y \perp\!\!\!\perp \bar{\sigma}_K \mid (\bar{L}_K, \bar{X}_K), \quad (2)$$

for all $k = 1, \ldots, K$. (Note, in addition certain positivity assumptions have to be satisfied which are detailed in Dawid & Didelez (2005)). Condition (2) is called *simple stability* as it implies that the covariates and the outcome are independent of how $X_k$ is generated once everything observable up to time $k$ is given. Less intuitive but weaker conditions than (2) that are sufficient for the identification of the intervention distribution as (1) are

$$Y \perp\!\!\!\perp \sigma_k \mid (\bar{X}_k, \bar{L}_k; \bar{\sigma}_{k-1} = \emptyset, \bar{\sigma}^{k+1} = \bar{s}^{k+1}) \qquad (3)$$

for all $k = 1, \ldots, K$, plus some additional more technical assumptions for which we refer to Dawid & Didelez (2005). A graphical check to see whether (3) is satisfied is given in Dawid & Didelez (2005) and is slightly different from Pearl & Robins (1995) as we allow for conditional interventions. Further Dawid & Didelez (2005) give conditions for (2) and (3) to be equivalent.

**Example 1:** We assume that the conditional independencies in the graph of Figure 1 hold, where $X_1 = X$, $X_2 = Z$, $L_1 = \emptyset$ and $L_2 = V$. Hence condition (2) for identifiability of the sequential treatment effect is satisfied if we observe $V$. If both interventions are atomic the intervention distribution can be calculated as

$$p(y; \sigma_X = s_x, \sigma_Z = s_z) = \sum_v p(y|x, z, v) p(v|x). \quad (4)$$

In case that the second intervention is conditional on $V$, specified by a function $a(v)$, we have

$$p(y; \sigma_X = s_x, \sigma_Z = s_{a(V)}) = \sum_v p(y|x, a(v), v) p(v|x).$$

And in case of a random intervention we have

$$p(y; \sigma_X = s_x, \sigma_Z = d_V) \\ = \sum_{z,v} p(y|x, z, v) \tilde{p}(z|v) p(v|x),$$

where $\tilde{p}(z|v)$ is the conditional distribution that $Z$ is drawn from under $\sigma_Z = d_V$.

## 3 DIRECT / INDIRECT EFFECTS

We now consider different notions of direct (indirect) effects of $X$ on $Y$ in the presence of a mediating variable $Z$. Roughly speaking, the direct effect is meant to reflect the influence of $X$ on $Y$ that is not conveyed by $Z$ and the indirect effect is the difference between total and direct effect. We use the notation $(X, Z, Y)$ from now on, instead of $(X_1, X_2, Y)$ from above, in order to facilitate comparison with Pearl (2001). It is assumed that $Z$ is affected by an intervention in $X$, i.e. $Z \not\!\perp\!\!\!\perp \sigma_X$ or, graphically, $Z$ should be a descendant of $X$. One might further demand that $Z$ consists of *all* intermediate variables between $X$ and $Y$ as in Pearl (2001), which would for instance not be the case in Figure 1. However, for the moment we prefer to speak of the direct effect of $X$ in relation to a particular choice $Z$, meaning the effect of $X$ that is not mediated by $Z$ but possibly mediated by other variables, and will come back to this question later.

In general one can say that the effect of $X$ not mediated by $Z$ is the effect of a change in $X$ while $Z$ is kept constant in some sense. The latter means that $Z$ is forced to arise in the same way while $X$ is being changed. Let $s \in \mathcal{S}_Z$, be a regime describing such an intervention in $Z$. In order for this not to depend on changes in $X$ we assume

$$Z \perp\!\!\!\perp \sigma_X | \sigma_Z = s. \qquad (5)$$

Graphically this means that under $\sigma_Z = s$ all arrows entering $Z$ coming from $X$ or its descendants are cut off. Then the *effect of $X$ not mediated by $Z$ under regime $s$* is defined as

$$DE_s(x, x^*; Y, Z) := E(Y; \sigma_X = s_x, \sigma_Z = s) \\ - E(Y; \sigma_X = s_{x^*}, \sigma_Z = s), \quad (6)$$

For brevity we refer to the above as 'direct effect' of $X$ under $s$ when the choice of $Z$ is clear. The basic ingredient in (6) is the intervention distribution $p(y; \sigma_X, \sigma_Z)$. Therefore it can be regarded as the effect of a sequential intervention where we first intervene in $X$, setting it to some value, and then we intervene in $Z$, drawing it from some distribution. The above conditions (2) or (3) with $X_1 = X$ and $X_2 = Z$ will ensure that the intervention distribution can be written as (1) for suitable $L_1, L_2$. Note that if in (6) we chose $s = \emptyset$ we would obtain the $ACE$, but this would typically not satisfy (5). For other choices of $s$ we obtain the following special cases.

### 3.1 CONTROLLED DIRECT EFFECT

The *average direct effect controlling for $Z = z$* is given by choosing $s = s_z$ in definition (6) yielding

$$CDE_z(x, x^*; Y, Z) := E(Y; \sigma_X = s_x, \sigma_Z = s_z) \\ - E(Y; \sigma_X = s_{x^*}, \sigma_Z = s_z), (7)$$

i.e. the expected difference in $Y$ between setting $X$ to $x$ as compared to $x^*$ while holding $Z$ constant at $z$. As $\sigma_Z = s_z$ cuts off all arrows entering $Z$ (5) is satisfied. The $CDE$ can be identified from observational data using (1) provided that (2) or (3) are satisfied.

It is easy to see that the quantity given by (7) can be perfectly meaningful even if $Z$ does not consist of all intermediate variables between $X$ and $Y$ as in Figure 1 for example. Here, the $CDE_z$ means mediated by $V$ but not by $Z$.

Note also that even if $Z$ is not affected by an intervention in $X$, i.e. if $Z \perp\!\!\!\perp \sigma_X$, the $CDE_z$ may depend on the value $z$, in particular if $X$ and $Z$ interact in their effect on $Y$. This somewhat contradicts the intuition that a direct effect should stay the same if $Z$ is not actually a mediating variable. For this and the above reason we think that it does not make much sense to define an indirect effect as difference between the total effect and the controlled direct effect.

## 3.2 STANDARDIZED DIRECT EFFECT

Following an idea of Geneletti (2005, 2006) we can choose the intervention $s$ in (6) to be a random, possibly conditional intervention, to obtain a standardized direct effect. This random intervention will be denoted by $d_W$ where $W$ is a subset of the parents of $Z$ except $X$ itself or any descendants of $X$, i.e. $W \subset \text{pa}_Z \backslash \{X \cup \text{de}_X\}$ — without referring to a graph this means $W \perp\!\!\!\perp (\sigma_X, \sigma_Z)$. Hence

$$p(z|\text{pa}_Z; \sigma_Z = d_w) := \tilde{p}(z|w),$$

where $\tilde{p}(z|w)$ is assumed to be a known distribution. The requirement $W \subset \text{pa}_Z \backslash \{X \cup \text{de}_X\}$ is necessary to satisfy (5); we want the intervention in $Z$ not to depend on $X$ so that we can indeed claim that $Z$ is being kept 'constant', i.e. generated from a distribution that remains the same while changing $X$.

The *average direct effect standardized w.r.t $Z$* is then defined as

$$SDE_{d_W}(x, x^*; Y, Z) := E(Y; \sigma_X = s_x, \sigma_Z = d_W) \\ - E(Y; \sigma_X = s_{x^*}, \sigma_Z = d_W).$$

The principle of standardization is well known from the comparison of mortality rates. This can be done in two ways, the age distribution of one of the two places or a historical composition can be used. The chosen age distribution would correspond to $\tilde{p}(z|w)$ and $X$ would be the places to be compared.

If we regard $X = X_1$ and $Z = X_2$ as sequential treatments like in Section 2.4, $W$ has to be a subset of $(L_1 \cup L_2) \cap \text{nd}_X$ and provided that (3) is satisfied we obtain $SDE_{d_W}(x, x^*; Y)$ using (1) as

$$\sum_{l_1, l_2, z} [E(Y|x, z, l_1, l_2)p(l_2|x, l_1) \\ - E(Y|x^*, z, l_1, l_2)p(l_2|x^*, l_1)]\tilde{p}(z|w)p(l_1). \quad (8)$$

**Example 2:** In Figure 2 the conditions (3) are satisfied for $X_1 = X, X_2 = Z, L_1 = \emptyset$ and $L_2 = W$.

## 3.3 NATURAL DIRECT EFFECT

The natural direct effect, in words, can be described as the effect of changing $X$ while $Z$ still arises randomly from its conditional distribution given $X$ set to the same baseline value throughout. In other words, we need to think of a way of intervening in $X$, changing it around, while $Z$ can still be generated as if $X$ was kept at its baseline value. For the natural direct effect to be empirically meaningful it needs to be made clear how such interventions can be carried out in practice. One might think for instance of some clever study designs, e.g. a placebo–type study (cf. Section 4.1).

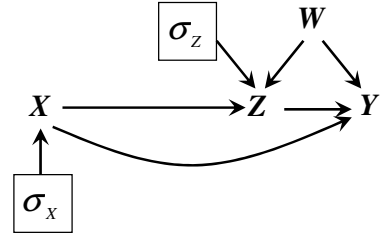

Figure 2: Influence diagram for Example 2.

The natural direct effect can be formulated as a special case of a standardized direct effect where $\tilde{p}(z|w)$ is chosen to be $p(z|w; \sigma_X = s_{x^*}, \sigma_Z = \emptyset)$, the conditional distribution of $Z$ within strata of $W$ where $X$ is being set to its baseline value. We symbolize this kind of intervention in $Z$ by $\sigma_Z = d_{W,x^*}$ meaning that

$$p(z|\text{pa}_Z; \sigma_Z = d_{w,x^*}) := p(z|w; \sigma_X = s_{x^*}, \sigma_Z = \emptyset).$$

A particular challenge is that the latter distribution is typically not known even if such an intervention can be carried out in practice. It could instead be estimated in a controlled study with $X$ fixed at its baseline value and the distribution of $Z$ measured within strata of $W$ as addressed in Section 4.

However, as opposed to the standardized direct effect the natural direct effect is only defined under an additional condition which restricts the choice of $W$ or $Z$. Consider the intervention distribution $p(y; \sigma_X = s_x, \sigma_Z = d_{W,x})$, where $X$ is set to the same value $x$ in both interventions. We want this to be the same as $p(y; \sigma_X = s_x)$ reflecting that if $X$ is set to $x$ and $Z$ is made to arise from its conditional distribution given $X$ set to $x$ this should just be like setting $X = x$ overall. In order for this to be fulfilled it is sufficient to assume

$$Y \perp\!\!\!\perp \sigma_Z \mid (Z, W; \sigma_X = s_x), \quad (9)$$

(and the same when substituting $\sigma_X = s_{x^*}$). The above needs to hold in addition to $W \perp\!\!\!\perp (\sigma_X, \sigma_Z)$ as required in Section 3.2 and by (5). Note that due to the definition of $\sigma_X = s_x$, (9) is the same as $Y \perp\!\!\!\perp \sigma_Z \mid (Z, W, X; \sigma_X = s_x)$. We then have

$$p(y; \sigma_X = s_x, \sigma_Z = d_{W,x}) =$$

$$\sum_{z,w} p(y|z,w;\sigma_X = s_x, \sigma_Z = d_{w,x})$$
$$p(z|w;\sigma_Z = d_{w,x})p(w)$$
$$= \sum_{z,w} p(y|z,w;\sigma_X = s_x, \sigma_Z = d_{w,x})$$
$$p(z|w;\sigma_X = s_x)p(w),$$

which is $p(y;\sigma_X = s_x)$ provided $p(y|z,w;\sigma_X = s_x, \sigma_Z = d_{w,x}) = p(y|z,w;\sigma_X = s_x, \sigma_Z = \emptyset)$ as implied by (9). This can be checked graphically: every back–door path from $Z$ to $Y$ that is not blocked by $X$ must be blocked by $W$ in the graph where all arrows entering $X$ have been deleted. This is equivalent to condition (11) of Pearl (2001).

The *average natural direct effect of $X$ on $Y$ w.r.t. $Z$* is now given by

$$NDE(x,x^*;Y,Z) = E(Y;\sigma_X = s_x, \sigma_Z = d_{W,x^*})$$
$$-E(Y;\sigma_X = s_{x^*}, \sigma_Z = d_{W,x^*}),$$

i.e. the expected difference in $Y$ between setting $X$ to $x$ as compared to $x^*$ while $Z$ arises randomly from its distribution with $X$ set to $x^*$ both times.

**Example 1 ctd.:** In Figure 1 condition (9) is not satisfied for $Z$ and $W = \emptyset$ as $V$ is a descendant of $X$ that affects $Y$ and $Z$. In our framework the $NDE$ of $X$ w.r.t. $Z$ is not defined for such a case as the intervention in $Z$ cannot sensibly be defined. It cannot be defined because there is an ambiguity of whether the direct effect of $X$ starts before or after $V$, i.e. whether $V$ should be conditional on $\sigma_X = x$ or $\sigma_X = x^*$. One can regard this as the 'wrong' choice of mediating variable, and alternatively choose $(V,Z)$ as mediating variables with regime $\sigma_{V,Z} = d_{x^*}$. However, to decide about a sensible approach in practice, one has to think about which interventions are actually feasible and therefore meaningful: can $V$ and $Z$ both be controlled to be generated from their distribution with $X$ set to $x^*$ while $X$ is actually changed from $x$ to $x^*$; or can we only control $Z$ to be generated from a distribution that is independent of changes in $X$ while $V$ is still affected by these changes. The latter case would correspond to the standardized direct effect, calculated similar to (8).

**Example 3:** For the example shown in Figure 3 we see that condition (9) is satisfied for $W = \emptyset$ and the natural direct effect of $X$ on $Y$ w.r.t. $Z$ is well defined as the effect not mediated by $Z$ but still mediated by $V$. It is hence not necessary to demand that $Z$ contain *all* intermediate variables.

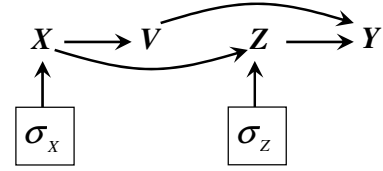

Figure 3: Influence diagram for Example 3.

### 3.4 NATURAL INDIRECT EFFECT

With (9) we have that $p(y;\sigma_X = s_x, \sigma_Z = d_{w,x}) = p(y;\sigma_X = s_x)$ and $p(y;\sigma_X = s_{x^*}, \sigma_Z = d_{w,x^*}) = p(y;\sigma_X = s_{x^*})$. This leads to the following definition of the *average natural indirect effect w.r.t. $Z$* as the difference between the total and the direct effect, $NIE = ACE - NDE$ yielding

$$NIE(x,x^*;Y,Z) = E(Y;\sigma_X = s_x, \sigma_Z = d_{w,x})$$
$$-E(Y;\sigma_X = s_x, \sigma_Z = d_{w,x^*}).$$

It can be interpreted as the effect of keeping $X$ fixed at $x$ while the conditional distribution of $Z$ is changed to arise conditionally on setting $X$ to $x$ and then $x^*$. Again, we would like to emphasize that the natural indirect effect is only empirically meaningful if one can at least in principle think of actual interventions that would accomplish this.

In summary, the identifiability of both the (natural) direct and indirect effects is essentially linked to the identifiability of $E(Y;\sigma_X = s_x, \sigma_Z = d_{w,x^*})$, $x \neq x^*$, which can be calculated from $p(y;\sigma_X = s_x, \sigma_Z = d_{x^*})$.

## 4 IDENTIFICATION

The controlled direct effect and the natural direct effect are both special cases of the standardized direct effect which in turn is a special case of sequential interventions. These can be identified by (2) or (3) if $p(z|\text{pa}_Z;\sigma_Z = s)$ is known. However, the natural direct effect poses the additional challenge that $p(z|\text{pa}_Z;\sigma_Z = d_{w,x^*})$ may be unknown. In the following we will pay special attention to the identification and estimation of the natural (in)direct effects.

### 4.1 STUDY DESIGN

In rare situations it may be possible to devise clever study designs where $X$ is randomized while $Z$ arises from its natural distribution under $X$ being set to a fixed value. An example is a clinical trial using placebos. Every subject receives a tablet, so that the psychological effect of thinking that one is being treated is the same, while some receive an active ingredient and others don't. Such a study design allows estimation of the direct, chemical/biological, effect of the active ingredient not mediated by any psychological effect. Note however that if the active ingredient has a noticeable side effect it is not guaranteed anymore that the psychology of treated and untreated subjects is comparable. Similar to Figure 1, with $V$ the side effect and $Z$ the 'psychology', condition (9) will be violated.

Another example is a study about managers' decisions on promotion based on CV's from men and women where the information about sex of the applicant has been changed, e.g. to everyone being male. The CV's represent the qualification arising naturally for men and women while the prejudice that managers may have is manipulated by changing the corresponding information on the CV. Such a study design would allow to estimate the indirect effect of sex through qualification on employment.

## 4.2 EXPERIMENTAL IDENTIFICATION

Experimental identification (Pearl, 2001) addresses the question whether the (in)direct effects can be estimated from randomized trials: one where $X$ and $Z$ are randomized and the effect on $Y$ is measured within strata of $W$; as this does not carry information on how $Z$ depends on $X$ we need a second experiment, where $X$ is randomized and its effect on $Z$ is measured. The condition needed for identification is a variation of (9) applied such that atomic interventions in $Z$ are covered.

**Example 2 ctd.:** In Figure 2, if we assume that the same graph is valid for $\sigma_Z$ taking values $s_z$ and $d_{w,x}$ condition (9) is satisfied. More precisely, we assume

$$p(y|w,z;\sigma_X = s_x, \sigma_Z = d_{w,x^*}) = p(y|w;\sigma_X = s_x, \sigma_Z = s_z),$$

which can be regarded as counterpart to condition (7) in Theorem 1 of Pearl (2001). The intervention distribution $p(y;\sigma_X = s_x, \sigma_Z = d_{W,x^*})$ is then given as

$$\sum_{z,w} p(y|w;\sigma_X = s_x, \sigma_Z = s_z)p(z|w;\sigma_X = s_{x^*})p(w)$$

and all terms may be estimated from two randomized studies as described above. Note that because all distributions as well as the data are gathered under $\sigma_X = s_x$ or $\sigma_X = s_{x^*}$, we do not need to worry about 'confounders' between $X$ and $Y$ or $X$ and $Z$.

## 4.3 NO–INTERACTION ASSUMPTION

As pointed out by Robins (2003), the natural (in)direct effects can be identified under the same conditions as, and will be equal to, the controlled direct effect if we assume that $X$ and $Z$ do not interact in their effect on $Y$. The no–interaction assumption implies that the direct effect $DE_s(x, x^*; Y, Z)$ from (6) is a function of $x$ and $x^*$ that does not depend on $s$, i.e. the way how $Z$ is manipulated, especially which value $Z$ is set to. An example is $E(Y; \sigma_X = s_x, \sigma_Z = s_z)$ being an additive function in $x$ and $z$.

However, the no–interaction assumption is likely to be violated in many practical applications. Some less restrictive parametrical assumptions that may allow identification of the (in)direct effects are discussed in Robins (2003) and TenHave et al. (2005).

## 4.4 OBSERVATIONAL STUDIES

If $p(z|w; \sigma_X = s_{x^*})$ is known the (in)direct effects can be identified from observational data with the conditions that allow the identification of sequential conditional interventions (3). If it is not known we must estimate this distribution. As this is the intervention distribution of $Z$ setting $X$ within strata of $W$, conditions for its identifiability are those for the $ACE$, e.g. the back–door criterion.

More precisely, we assume there exist possibly overlapping sets of covariates $S, L_1, L_2$ and $W \subset (L_1 \cup L_2)$ such that $W$ and $Z$ together satisfy (9) and

1. $(W, S, L_1) \perp\!\!\!\perp (\sigma_X, \sigma_Z)$, i.e. they are non descendants of $X$ and $Z$;

2. $L_2 \perp\!\!\!\perp \sigma_Z$, i.e. $L_2$ can be a descendant of $X$ but not of $Z$;

3. $Y \perp\!\!\!\perp \sigma_X | (X, W, L_1; \sigma_Z = d_{W,x^*})$;

4. $Y \perp\!\!\!\perp \sigma_Z | (X, Z, L_1, L_2; \sigma_X = \emptyset)$;

5. $Z \perp\!\!\!\perp \sigma_X | (X, W, S; \sigma_Z = \emptyset)$.

In the above, the conditions on $(L_1, L_2)$, 3. and 4., are like for the sequential treatment setting (3). The conditions on $S$ are like the back-door criterion for the effect of $X$ on $Z$.

We now show how the above allow the identification the natural (in)direct effects, or more precisely the required intervention distribution $p(y; \sigma_X = s_x, \sigma_Z = d_{W,x^*})$. With assumption 1. we have

$$p(y; \sigma_X = s_x, \sigma_Z = d_{W,x^*})$$
$$= \sum_{w,l_1} p(y|w, l_1; \sigma_X = s_x, \sigma_Z = d_{w,x^*})p(w, l_1)$$

(where $\sigma_X = \emptyset, \sigma_Z = \emptyset$ have been omitted) and hence with 3. it follows that the above is equal to

$$\sum_{l_1} p(y|w, l_1, x; \sigma_X = \emptyset, \sigma_Z = d_{w,x^*})p(w, l_1).$$

Now, by definition we have

$$p(y|w, l_1, x; \sigma_Z = d_{w,x^*}) = \sum_{l_2} p(y|l_1, l_2, x; \sigma_Z = d_{w,x^*}) p(l_2|w, l_1, x; \sigma_Z = d_{w,x^*})$$

and assumptions 1. and 2. yield that this is equal to

$$\sum_{l_2} p(y|l_1, l_2, x; \sigma_Z = d_{w,x^*}) p(l_2|w, l_1, x).$$

Further,

$$p(y|l_1, l_2, x; \sigma_Z = d_{w,x^*}) = \sum_z p(y|l_1, l_2, z, x; \sigma_Z = d_{w,x^*}) \times p(z|l_1, l_2, x; \sigma_Z = d_{w,x^*})$$

which by definition of $\sigma_Z = d_{w,x^*}$ and assumption 4. is

$$\sum_z p(y|l_1, l_2, z, x) p(z|w; \sigma_X = s_{x^*}).$$

Now we just need that

$$p(z|w; \sigma_X = s_{x^*}) = \sum_s p(z|w, s, x^*) p(s|w)$$

which it is by assumptions 1. and 5. using the back door argument within strata of $W$.

Hence the intervention distribution $p(y; \sigma_X = s_x, \sigma_Z = d_{W,x^*})$ is given by

$$\sum_{s, l_1, l_2, z} p(y|l_1, l_2, z, x) p(z|w, s, x^*) p(l_2|x, l_1) p(s|w) p(l_1),$$

where all distributions are observational, i.e. conditional on $\sigma_X = \emptyset$ and $\sigma_Z = \emptyset$ and can thus be estimated from observational data on $X, Y, Z, S, L_1, L_2$.

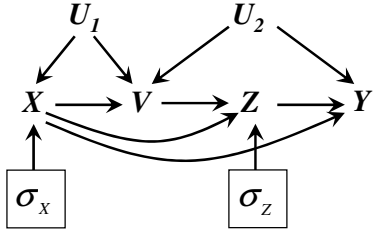

Figure 4: Influence diagram for Example 4.

**Example 4:** Assume the influence diagram given in Figure 4. This is an example modified from Pearl & Robins (1995). It is shown in Dawid & Didelez (2005) that if $(U_1, U_2)$ are unobservable only unconditional interventions in $Z$ can be identified with the rules (3), because $Y \perp\!\!\!\perp \sigma_X | (X; \sigma_Z = s)$ holds only if $\sigma_Z = s$ 'cuts off' the arrow from $V$ to $Z$ as shown in Figure 5 and as is assumed in Pearl & Robins (1995). However, $V$ cannot be ignored altogether as we need $Y \perp\!\!\!\perp \sigma_Z | (X, Z, V; \sigma_X = \emptyset)$. The corresponding intervention distribution is given by the same formula as (4). Consequently the controlled direct effect as well as the standardized direct effect (with an unconditional random intervention in $Z$) can be estimated in the same way, the latter based on

$$p(y; \sigma_X = s_x, \sigma_Z = d) = \sum_{z,v} p(y|x, z, v) \tilde{p}(z) p(v|x),$$

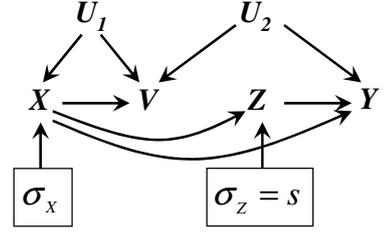

Figure 5: Modified influence diagram for Example 4 with unconditional intervention in $Z$.

In order to estimate the natural direct effect of $X$ we first need to check whether a variable $W$ satisfying (9) exists so that the direct effect is well defined. This is only the case for the choice $W = U_2$. For the remaining conditions, we have to choose $L_1 \cup L_2 = W = U_2$ and $S = U_1$. Note that condition 3. can be verified in the graph given in Figure 6 where the intervention $\sigma_Z = d_{U_2,x^*}$ makes $U_2$ a parent of $Z$ while $V$ is not a parent anymore. Thus, we obtain the intervention distribution required for the direct or indirect causal effect as

$$p(y; \sigma_X = s_x, \sigma_Z = d_{U_2,x^*}) = \sum_{u_1, u_2, z} p(y|x, z, u_2) p(z|u_1, u_2, x^*) p(u_1, u_2).$$

## 5 DISCUSSION

The present paper demonstrates how the notions of direct and indirect effects can be formulated in terms of interventions and hence can be regarded as a special case of sequential treatments yielding conditions for identifiability from observational data. Our proposal generalizes the ideas of Pearl (2001) to standardized direct effects, the natural direct effect being a special case. The latter involves interventions that are particular in that they are random, possibly drawn from an unknown distribution, which requires some additional conditions for identifiability. The concepts we have introduced assume an underlying DAG relating the variables of interest and other variables. Further analysis is desirable to clarify, and where possible eliminate, sensitivity to such assumptions.

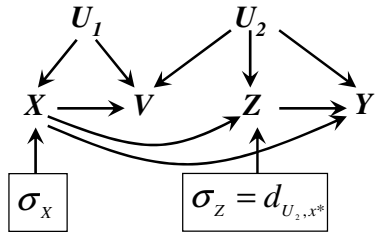

Figure 6: Modified influence diagram for Example 4 with intervention in $Z$ to reflect direct effect of $X$.

Unlike Pearl (2001) we do not require the mediating variable $Z$ to consist of all intermediate variables, instead we consider 'direct effect' as meaning 'not mediated by $Z$'. Moreover, we believe if we took seriously that $Z$ has to contain all intermediate variables the direct effect of $X$ would always be zero. Hence such a requirement does not seem to make sense suggesting that 'the' direct effect is misleading and should be replaced by 'direct effect in relation to $Z$'.

In our view, it is fruitful to insist on a formulation of causal quantities in general, and of (in)direct effects in particular, in terms of interventions. It forces us to think about the empirical content of these quantities. As seen in Section 3.3 these considerations are relevant for a sensible and meaningful choice of the mediating variable $Z$ and hence for the data to be collected. Robins (2003) calls the natural (in)direct effects non–manipulative parameters claiming that they cannot be put in terms of interventions or manipulations; we show that such a formulation is possible but agree that only in rare cases can one think of actual manipulations that capture the intuitive notion behind natural (in)direct effects, such as given in Section 4.1. This suggests that often it is not so much the natural (in)direct effects that are of practical interest but effects where the distribution of $Z$ is controlled in some different way. Thus, the more general notion of standardized direct effects seems more relevant.

### Acknowledgements

This work has been developed while Vanessa Didelez was member of an international research group studying Statistical Analysis of Complex Event History Data at the Centre for Advanced Study at the Norwegian Academy of Science and Letters in Oslo during the academic year 2005/2006.